\documentclass{article}

\usepackage{arxiv}

\usepackage[utf8]{inputenc} 
\usepackage[table]{xcolor}
\usepackage[T1]{fontenc}    
\usepackage[binary-units = true]{siunitx}
\usepackage{url}            
\usepackage{booktabs}       
\usepackage{amsfonts}       
\usepackage{amsmath}        
\usepackage{nicefrac}       
\usepackage{microtype}      
\usepackage{lipsum}         
\usepackage{graphicx}       
\usepackage{verbatim}       
\usepackage{svg}            
\usepackage{authblk}        
\usepackage{bm}
\usepackage{wrapfig} 
\usepackage[figurename=Fig.]{caption}
\usepackage[frozencache=true,cachedir=.]{minted}

\usepackage{array}
\usepackage{pifont}

\graphicspath{ {./images/} }
\usepackage{hyperref}   

\title{aDWI-BIDS: an extension to the brain imaging data structure for advanced diffusion weighted imaging}

\author[1,2]{James Gholam}
\author[3]{Filip Szczepankiewicz}
\author[1,2,4]{Chantal M.W. Tax}
\author[2,5]{Lars Mueller}
\author[2,5]{Emre Kopanoglu} 
\author[3]{Markus Nilsson} 
\author[6]{Santiago Aja-Fernandez}
\author[1]{Matt Griffin} 
\author[2,5]{Derek K. Jones}
\author[1,2]{Leandro Beltrachini}

\affil[1]{School of Physics and Astronomy, Cardiff University, Cardiff, United Kingdom}
\affil[2]{Cardiff Univeristy Brain Research Imaging Centre (CUBRIC), Cardiff, United Kingdom}
\affil[3]{Department of Diagnostic Radiology, Lund University, Lund, Sweden}
\affil[4]{Image Sciences Institute, University Medical Center Utrecht, Utrecht, Netherlands}
\affil[5]{School of Psychology, Cardiff University, Cardiff, United Kingdom}
\affil[6]{Universidad de Valladolid, Valladolid, Spain}

\begin{document}
\maketitle
\begin{abstract}
Diffusion weighted imaging techniques permit us to infer microstructural detail in biological tissue in vivo and noninvasively. Modern sequences are based on advanced diffusion encoding schemes, allowing probing of more revealing measures of tissue microstructure than the standard apparent diffusion coefficient or fractional anisotropy. Though these methods may result in faster or more revealing acquisitions, they generally demand prior knowledge of sequence-specific parameters for which there is no accepted sharing standard. Here, we present a metadata labelling scheme suitable for the needs of developers and users within the diffusion neuroimaging community alike: a lightweight, unambiguous parametric map relaying acqusition parameters. This extensible scheme supports a wide spectrum of diffusion encoding methods, from single diffusion encoding to highly complex sequences involving arbitrary gradient waveforms. Built under the brain imaging data structure (BIDS), it allows storage of advanced diffusion MRI data comprehensively alongside any other neuroimaging information, facilitating processing pipelines and multimodal analyses. We illustrate the usefulness of this BIDS-extension with a range of example data, and discuss the extension's impact on pre- and post-processing software. 
\end{abstract}

\section{Introduction}
Diffusion weighted imaging (DWI) has evolved substantially over its lifetime, encompassing a range of related techniques seeking to better resolve and image diffusion processes in vivo. New sequences and sampling methods now permit resolution of details beyond the determination of the diffusion tensor. Modelling approaches like CHARMED~\cite{assaf_composite_2005} and NODDI~\cite{zhang_noddi_2012} have motivated MRI protocols with the power to efficiently resolve major microstructural properties across a wide range of tissues. Novel techniques require arbitrary gradient waveform trajectories for multiple purposes, including tensor-valued diffusion encoding, more specific feature encoding (such as flow or exchange), as well as for combatting certain artifacts (e.g. Maxwell, background field effects etc.). As well, compared to single diffusion encoding, double~\cite{cory_measurement_1990} and multiple diffusion encoding methods (SDE, DDE and MDE respectively~\cite{shemesh_conventions_2016}) offer a way to disentangle several microscopic diffusion phenomena that may be collated in an SDE experiment~\cite{ozarslan_microscopic_2008, callaghan_komlosh_2002}, resolve diffusion at small time scales~\cite{ozarslan_mr_2007, clark_diffusion_2001, shemesh_measuring_2009}, amongst other possibilities offered by this technique.

Modern diffusion sensitised protocols may densely sample different orientations and b-values~\cite{descoteaux_hardi_2015, assaf_composite_2005, zhang_noddi_2012}, employ multiple gradient and RF pulses within diffusion preparation~\cite{shemesh_conventions_2016, avram_vivo_2013, topgaard_isotropic_2015, topgaard_multidimensional_2017}, utilise non-standard gradient pulses~\cite{westin_q-space_2016, sjolund_constrained_2015, descoteaux_hardi_2015, szczepankiewicz_quantification_2015}, or incorporate interleaving and parallel imaging to speedup acquisitions, resulting in parameter variation between measurements (either by volume or by slice~\cite{HUTTER_2018214, hutter_dynamic_2017}). Processing data from these approaches and their combinations require detailed information about the diffusion preparation used in each image or volume. This demands more information than is currently supplied by the DICOM files output by scanners. Limited standardisation of data formats results in significant use of ad-hoc methods for communicating sequence parameters~\cite{descoteaux_hardi_2015}. For those producing and consuming archive-quality data, such methods are too variable to be reasonably programmed for, obstructing large scale analysis inherent to population studies, pipeline comparison, and machine learning~\cite{gal_data_2019, data_harmon_Bittner2021}.

This limitation in DWI metadata is broadly overcome by the Brain Imaging Data Structure (BIDS)\cite{gorgolewski_brain_2016}, an extensible storage standard for neuroimaging data. BIDS presents data in a consistent format (\path{.nii} and \path{.nii.gz} files), permitting substantial code reuse. Image data from an acquisition is supplemented by metadata in \path{.json} files, drawn directly from parent DICOM files during conversion~\cite{lopez-novoa_bids_2019}. BIDS specifies a coherent file naming scheme, as well as a directory structure that divides acquisitions on the basis of subject, session, and imaging modality. A large range of neuroimaging data are now fully or partially supported by BIDS, incorporating structural and functional MRI~\cite{gorgolewski_brain_2016}, positron emission tomography~\cite{BIDS_PET}, and electro/magnetoencephalography~\cite{pernet_eeg-bids_2019, niso_meg-bids_2018}, as well as several calibration standards. The logical structure of BIDS simplifies traversing datasets, allowing similar tools and libraries to access dissimilar data. BIDS gives several methods to extend its functionality, from adding new keys to metadata, to the addition of entire supplementary files. 

Though numerous advantages are provided by the BIDS, it does not include a standardisation for the aforementioned advanced DWI methodologies, being limited to the most basic diffusion sequences. To solve this issue, we introduce aDWI-BIDS, an extension to the brain imaging data structure for advanced diffusion weighted imaging. This extension offers a scheme to communicate complex diffusion encoding methods and to unambiguously relate them to data acquired. By introducing the concept of sequence events, we provide a modular approach to the description of arbitrary DWI sequences, including gradients and RF pulses. Moreover, the methodology allows specification of sequence parameters at different scales, ranging from the entire field of view of a time series, to the single voxel level. We illustrate the benefits of the extension to describe diverse DWI data, as well as to regularise the corresponding processing pipelines. 

This paper is organised as follows. In Section~\ref{sec:methods} we detail the extension proposed. After providing a justification for its need (Section~\ref{sec:need_ext}), we introduce the files that allow a proper description of the experimental sequence employed, detailing each of them in the corresponding subsections. Moreover, in Section~\ref{sec:indr_perturb} we explain the concepts of perturbations and indirections, which are shown to enhance the flexibility and clarity of the resulting files while minimising data redundancy. In Section~\ref{sec:illustrations} we present a set of illustrations that show the application of the standard to different use cases, from widely accepted standards to state-of-the-art designs. In addition, we show its adaptability to describe other data fields, such as RF and readout pulses. Finally, in Section~\ref{sec:discussion} we discuss the results and present perspectives of its utilisation.

\section{Methods} \label{sec:methods}
\subsection{The BIDS standard and the need for an extension}\label{sec:need_ext}

The Brain Imaging Data Structure (BIDS) is a modular, extensible representation for neuroimaging data, collecting and arranging multiple imaging modalities in a single directory. This directory collects data first by subject, then acquisition session, then modality (Fig.~\ref{fig:structure}). This means multiple (potentially simultaneously collected) data may be present in an acquisition comprised of several modalities and data types. New modalities may be added without disturbing this structure, all sharing certain common principles and file standards. The underlying motivation is to seamlessly integrate existing software where possible with minimal extension.

BIDS conveys MRI data specifically in three primary formats, with some modalities using additional files. DWI gives imaging data using the NIfTI format, as well as a \verb|.json| "sidecar" file. Diffusion directions and b-values are given using \verb|.bvec| and \verb|.bval| files. Acquisition-specific metadata not conveyed by a NIfTI is given in the sidecar. The \verb|.bvec| and \verb|.bval| files provide volume-by-volume variation in the acquisition parameters implied. This limited set of parameters (particularly those from the \verb|.bvec| and \verb|.bval| files) restricts the range of diffusion imaging modalities which may be fully conveyed by BIDS.

BIDS presently communicates DWI data, specified in terms of \verb|.bvec| and \verb|.bval| files, an explicit sequence name, and a collection of whole-dataset metadata~\cite{gorgolewski_brain_2016}. This has been satisfactory for many applications, and can be misused to reproduce a wide range of behaviour based on the reported data with the \verb|.bvec| and \verb|.bval|. Several cases expose limitations in this approach, e.g. in the case of sequences:

\begin{itemize}
    \item using pulse shapes other than trapezoids, e.g.,  'long-narrow'~\cite{bernstein_handbook_2004}, oscillating~\cite{baron_oscillating_2014}, free~\cite{szczepankiewicz_quantification_2015}, quiet~\cite{edelstein_making_2002} gradients, as well as numerous compensating waveforms;
    \item using multiple pulses where amplitude, duration or spacing of RF or gradient pulses represents relevant metadata, e.g. DDE~\cite{cory_measurement_1990}, FEXI~\cite{nilsson_noninvasive_2013}, STEAM~\cite{matthaei_steam_1986, merboldt_steam_1991} or diffusion-TSE~\cite{yoshida_image_2016};
    \item with different encoding from slice to slice~\cite{HUTTER_2018214, hutter_dynamic_2017}.
\end{itemize}

Sequences implementing any of these properties cannot be specified by BIDS in its current form, and no consistent method exists within BIDS to relate them. Furthermore, many of these approaches fundamentally change either the ordering or interpretation of data, limiting the adoption of BIDS in an experimental context. Microstructural imaging methods can offer otherwise absent detail about the properties of tissue, and may allow faster acquisition with the same apparatus. This motivates the generation of a single, standardised metadata labelling scheme to encompass these conditions.

\subsection{Extensible descriptions of diffusion encoding: the aDWI-BIDS}

\begin{figure}[h!]
    \centering
    \includegraphics[scale = .9, trim=5cm 5cm 5cm 5cm, clip]{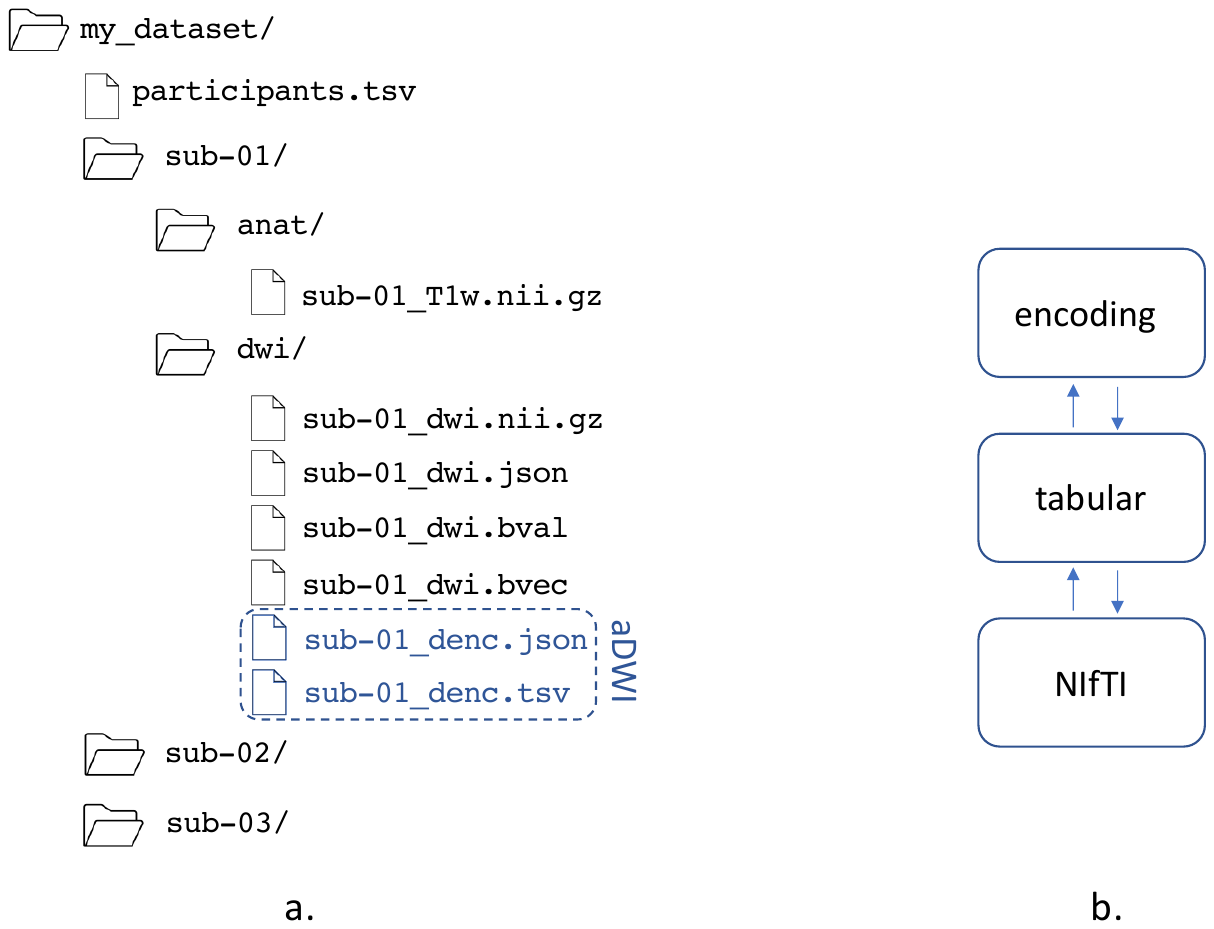}
    \caption{BIDS structure and the aDWI extension. a. Folder structure of BIDS, with the encoding and tabular files defined as part of the aDWI extension marked in blue. Additional files may be required for indirections (see Section~\ref{sec:indr_perturb}). 
    b. Schematic relation between the encoding, tabular, and NIfTI files, with the second orchestrating the data and metadata from the other two.}
    \label{fig:structure}
\end{figure}

The aDWI-BIDS extension is primarily comprised of a matched pair of documents in addition to the imaging (NIfTI or \verb|.nii|) data: an {\it encoding} (\verb|.json|) file comprising all the information related to the diffusion sequences employed throughout the acquisition, and a {\it tabular} (\verb|.tsv|) file relating the sequence events in the JSON to a particular subset of the NIfTI document. The interplay between these three files (\verb|.json|, \verb|.tsv|, and \verb|.nii|) is crucial for guaranteeing maximum flexibility of the standard, and consequently requires a detailed description (Fig.~\ref{fig:structure}). The generic structure of a linked \verb|.json| and \verb|.tsv| file is defined in the BIDS documentation\footnote{ \url{https://bids-specification.readthedocs.io/en/stable/02-common-principles.html}}. 

In broad terms, the files possess three distinct roles within aDWI-BIDS:
\begin{itemize}
    \item {\bf NIfTI (\verb|.nii|)}: Supplies voxel data arranged into slices, volumes or times series of either slices or volumes. Different slices or volumes may represent data collected with different experimental parameters (such as an interleaved acquisition).
    \item {\bf Encoding (\verb|.json|)}: Describes a set of collections of experimental sequence parameters used in one or more NIfTI files. Collections of parameters are given alongside indices to differentiate them.
    \item {\bf Tabular (\verb|.tsv|)}: Describes how the data in a single NIfTI was recorded. This is given for each slice or volume in a series by labelling each item with an index corresponding to one of the set defined in the encoding file. The tabular file may also define {\it perturbations} to parameters, representing slight variations in one or more parameter's value from slice to slice or volume to volume. This helps to limit redundant data.
\end{itemize}

The three files are named similarly to other BIDS diffusion data, using the following ordered list of tags within the filename:
\begin{minted}{text}
sub-<label>[_ses-<label>][_acq-<label>][_dir-<label>][_run-<index>]_dwi.nii[.gz]
sub-<label>[_ses-<label>][_acq-<label>][_dir-<label>][_run-<index>]_denc.tsv
sub-<label>[_ses-<label>][_acq-<label>][_dir-<label>][_run-<index>]_denc.json
\end{minted}
The rules of file naming in BIDS mean the first three tags (\verb|sub-<label>[_ses-<label>][_acq-<label>]|) do not pertain to the imaging modality. This facilitates the use of the inheritance principle, whereby the encoding and tabular files may omit some or all of these tags if they are placed higher in the directory structure. If this is the case, files in sub-directories "inherit" the parameters given by the tabular and encoding files, if they are of matched modality. The use of the \verb|[_dir-<label>]| tag is largely superseded by the tabular file, which specifies the one or more directions used in the acquisition.

\subsubsection{The tabular file (.tsv)}\label{sec:tabular}

Imaging data is stored in NIfTI files, in compliance with the BIDS format. NIfTI files are multidimensional by design, specifying up to seven dimensions~\cite{mjenkinson_nifti-1_nodate}. Data in a NIfTI may represent 2D slices, 3D volumes, or a series of either. This means a NIfTI may be flattened into a 1D structure, and components numbered using a parameter map (the tabular file). Each item in the series is labelled with incrementing index $t$ in the order in which the data occurs within the NIfTI. For multiband and other sequences, this may fragment volumes into an interleaved arrangement. Consequently, we reference which volume a slice belongs to with index $v$. To indicate the location of a slice in a volume, we label it with an index $k$ representing its position within a stack of slices in the NIfTI data space. This may be optionally extended to row or even voxel level description with more indices. It is expected that volume level description will be favoured for most applications. 

Additionally, each slice or volume may be labelled with zero or more perturbations. These may represent small variations in the parameters given, e.g., a sequence which varies pulse duration $\delta$ may provide unique values of $\delta$ for every slice or volume. Perturbations allow efficient reuse of parametric data with a minimum of repetition. More complex perturbations may require clear rules as to how an encoding file's parameters change as a result. A major class of perturbations are rotations and scaling of gradient activity. This may be provided by three Euler angles and a scaling coefficient per slice or volume, labelled $x$, $y$, $z$, and $s$, respectively. Euler rotations are in the order given, i.e., rotations around axes $x$, $y$ and $z$. A list of indices for slice-by-slice parametric variation are given in Table \ref{tab:tsv}, also showing the rotation and scaling perturbations. For convenience, the absence of a parameter indicates no change between all acquisitions.

\begin{table}[ht]
\begin{tabular}{|m{1cm}|m{3.5cm}|m{11cm}|}
\hline
Index & Name & Meaning \\ \hline
    \verb|t| &  Temporal slice index & Numbering index in order of acquisition\\ \hline
    \verb|v| &  Volume index & Index such that acquisitions with the same \verb|v| are of the same volume\\ \hline
    \verb|k| &  Geometric slice index &  Index incrementing alongside the slice number in the NIfTI file\\ \hline
    \verb|d| &  Diffusion encoding index & Index such that acquisitions with the same \verb|d| are encoded the same way \\ \hline
    \verb|x| &  Euler angle \textbf{i} & Euler rotation of sequence parameters about axis \verb|x|\\ \hline
    \verb|y| &  Euler angle \textbf{j} & Euler rotation of sequence parameters about axis \verb|y|\\ \hline
    \verb|z| &  Euler angle \textbf{k} & Euler rotation of sequence parameters about axis \verb|z|\\ \hline
    \verb|s| &  Scaling coefficient & Scaling of gradient amplitude by a linear scaling coefficient \\ \hline
\end{tabular}
\caption{Columns found in the tabular file. \texttt{t, v} and \texttt{k} refer the location of the data in the NIfTI (time, volume and slice location respectively). \texttt{d} gives the diffusion encoding object from the encoding file used by this acquisition. \texttt{x,y,z} and \texttt{s} are perturbations to the encoding object \texttt{d}}
\label{tab:tsv}
\end{table}

\subsubsection{The encoding file (.json)}\label{sec:encoding_file}
The encoding file is responsible for supplying sets of parameters varied in one or more NIfTI files. It does so by defining the column headers of the tabular file, and may specify how values under that header are interpreted. For example, if a header $d$ with values \verb|[0,1,2,3]| is used in the tabular file, we can interpret the meaning of these values by linking each of them in the encoding file:

\begin{minted}{json}
{"d": {"Levels": {
	"0": {"Object defining B0 reference"},
	"1": {"Object defining B1000 sPGSE" },
	"2": {"Object defining B2000 sPGSE" },
	"3": {"Object defining B3000 sPGSE" }}}
\end{minted}

where each "Level" in \verb|[0,1,2,3]| supplies a unique JSON object, describing a different diffusion encoding object. Each encoding object is a self-contained set of parameters that defines sequence activity according to a specified ontology. This is a deliberately unrestrictive system - if a sequence is parameterisable in a simple way, this should be possible as data stored within this object.

\begin{figure}[h!]
    \centering

    \includegraphics[width=0.7\textwidth]{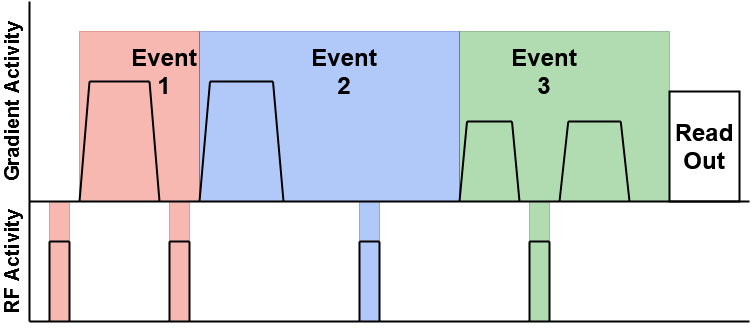}
    \caption{Diagrammatic outline of how sequence activity can be segmented into discrete events. Each coloured box represents a distinct object, or "event". Each event possesses a "temporal origin", which here lays at the left side of each coloured box within gradient activity. Each event is responsible for indicating when the next event will begin. All subevents are given relative to a time offset from that event's temporal origin. As in Event 1, this offset may be negative.}
    \label{fig:basic_events}
\end{figure}

We also outline a system to generically communicate any kind of sequence activity occurring as a set of distinct events in time. Activity may consequently be parameterised in terms of a domain's specific notation. If this is not possible, tools exist to give activity as an array of amplitudes spaced in time. Regions of time are divided into "events", as shown in Figure~\ref{fig:basic_events}. Events can describe RF, gradient, or readout activity (the components of which are called "subevents"). Events possess several attributes that permit this:
\begin{itemize}
    \item An implicit temporal origin that subevents occur relative to.
    \item A duration until the subsequent event begins (\verb|t_ev|).
    \item A metadata field, for properties true for the whole event (\verb|"meta":{}|).
    \item A field for explicit transformations of an event (\verb|"trf":{}|).
    \item A collection of uniquely named subevents belonging to the parent event.
    \item A unique "type" value that permits programmatic identification (\verb|"ev_type"|).
\end{itemize}

Beyond this, subevents are unfettered in their structure. Parameterised subevents may be structured to most simply convey their activity in terms of coefficients. This parameterised data is expanded prior to use according to the use case, potentially into derivative metrics, or into amplitudes on a temporal raster. When activity is too complex to specify parametrically, an explicit array of amplitudes may be given, with facility for dense binary encoding without reliance on base64 representation.

Certain types of events and subevents are very common, such as trapezoidal pulses, some types of paired pulse gradients, free-gradient waveforms, \textit{sinc} pulses and others. Some of these use cases have provisional object schemas defined, and will be described and illustrated in Section~\ref{sec:illustrations}. We note that schemas may be explicitly defined for JSON objects using the well supported JSON Schema framework~\cite{handrews-json-schema-02}. Within this framework, it is simple to iterate over the array of events, and identify how it is structured, and call appropriate code to process and expand these events.

\subsubsection{Perturbations and indirections}\label{sec:indr_perturb}
Data in aDWI-BIDS is optimised for density and reuseability, achieved partially by {\it perturbations} and {\it indirections}, respectively. Perturbations are a powerful feature describing data in terms of modifications to a prototypical sequence. In this, the encoding file describes the prototypical sequence, while the tabular file gives the corresponding variations. The tabular file is hence a kind of parameter map - a list of keys (properties) and their values (in a given region of acquisition), relating to a list of objects (the encoding file) that possesses these properties. In this sense, perturbations are simply transforms of a prototypical object. However two distinct forms are present.

The first is a functional transform (strictly, a \textit{functor}) of data within an event, such as rotation using Euler angles $x$, $y$ and $z$, or scaling with $s$. These modify one or more parameters in an object, mapping object $A$ to object $B$ according to a prescribed scheme. These transforms permit the tabular file to mutate the encoding object in complex ways. This is shown in Section~\ref{sec:ex2DDE}, where a double diffusion encoding sequence experiences a shifting $\Delta$ that requires the related RF pulses to move correspondingly.

The second kind of permutation is that of a direct modification to an event parameter. In this, a value for an event is substituted for the value of the permutation. This permits overwriting of parameters in a prototypical object in a tabular way. This can be used to track, e.g., an experiment with a pulse duration $\delta$ which varies between volumes, or seeing how two RF pulse shapes modify the sequence. In this instance, the desired parameter is simply given as a column header within the tabular file. The access pattern used to refer to a key within the encoding object is used as the name for the column header. Hence, a header \verb|[1]."gr_pair"."t_bdel"| acting on an encoding object of \verb|"ev_type":"SDE"| will overwrite the default parameter given for this object.

Indirections allow users to store binary data as a separate file, but to link that data to particular encoding objects. This is achieved by simply providing an indirection path to an event's \verb|"meta:{}"| field. Binary standards which are structural supersets of JSON permit detailed objects to be nested within an indirection. Using the CBOR~\cite{CBOR} standard permits regions of binary data to be structured as in JSON. We can then refer to objects in the indirected file using an appropriate object, in place of an array of data (e.g., we replace the gradient waveform array \verb|xgrad:[0,1,...,n]| with \verb|xgrad:{indr:b_xgrad}|, where \verb|b_xgrad| is the key of the object in the indirected file). By doing this, we increase readability of the encoding file by avoiding unnecessary cluttering. This is shown in detail in Section~\ref{sec:fwf}, where a free gradient sequence is shown to use this approach to compress data.

\subsubsection{Access patterns}
The structured data comprising aDWI-BIDS must be accessed in a characteristic pattern to correctly represent the underlying data. In many cases, the order of operations is unimportant, and we may freely access invariant parameters used in a sequence, or refer only to tabular variation in parameters without concerning ourselves with how this changes the underlying object. 

We refer to the combined information conveyed by the NIfTI, tabular and encoding files as the "expanded sequence". This object is implicitly an array, with as many entries as rows in a tabular file, and with each entry being an instance of an encoding object. We can view this structure as an array of class instances, with each instance possessing the same collection of methods. Certain methods must hence be implemented:
\begin{itemize}
    \item Object expansion to a set of values in time, e.g., a gradient amplitude array or RF phase / amplitude map.
    \item Methods to determine duration and amplitudes in an event.
    \item Rotations and scaling of the object (this may result in an empty method).
    \item Expansion of any indirections the object uses.
    \item Method to apply transformations to an object (one per transformation).
\end{itemize}

In general, before using an object, we recommend to i. apply any permutations of the second kind, ii. expand any indirections possessed, and then iii. apply permutations of the first kind (e.g., transformations). This may be done with varying efficiency. To minimise memory use and operations, the following access pattern may be adopted:

\begin{enumerate}

    \item Check the tabular file for permutations of the second kind. Unless these are indirections, do nothing. If these are indirections, go straight to step 3.
    \item Determine all indirections which will be used and their targets. Open the relevant files and load these to memory.
    \item Working row-by-row, substitute data from indirections wherever the file is referenced and apply all remaining permutations (of either type).
\end{enumerate}

As the data is accessed in each row, an expanded encoding object is instantiated to be handled, or have other methods called upon it for a user's application.

\section{Illustrations}\label{sec:illustrations}

In the following subsections, we show how aDWI-BIDS may be used to convey highly descriptive sequence behaviour for several types of diffusion encodings. We show how the encoding file will appear for each case, using one or more encoding sub-objects. Moreover, we demonstrate how the tabular file may be used alongside the encoding file to permute its interpretation, representing a set of parameters, some of them varying between acquisitions.

\subsection{Example 1: single diffusion encoding}\label{sec:ex1stej}

In the prototypical Single Diffusion Encoding (SDE; also known as Stejskal-Tanner or single Pulse Gradient Spin Echo) experiment, two RF pulses and two gradient pulses sensitise a sequence to molecular diffusion. A 90$^{\circ}$ RF storage pulse establishes a partial phase coherence in a system's spin precession in a strong magnetic field. A short gradient pulse dephases spins, and the system is permitted to evolve (and diffuse) for a period of time. During this evolution, a 180$^{\circ}$ refocusing RF pulse reverses the system's phase evolution. A final gradient pulse, identical to the first refocuses the system after another period passes, and readout can determine the spatial variation of magnetisation in the system~\cite{Stejskal}. 

\begin{figure}[ht]
    \centering
    \includegraphics[scale = .85, trim=5cm 0.5cm 5cm 4cm, clip]{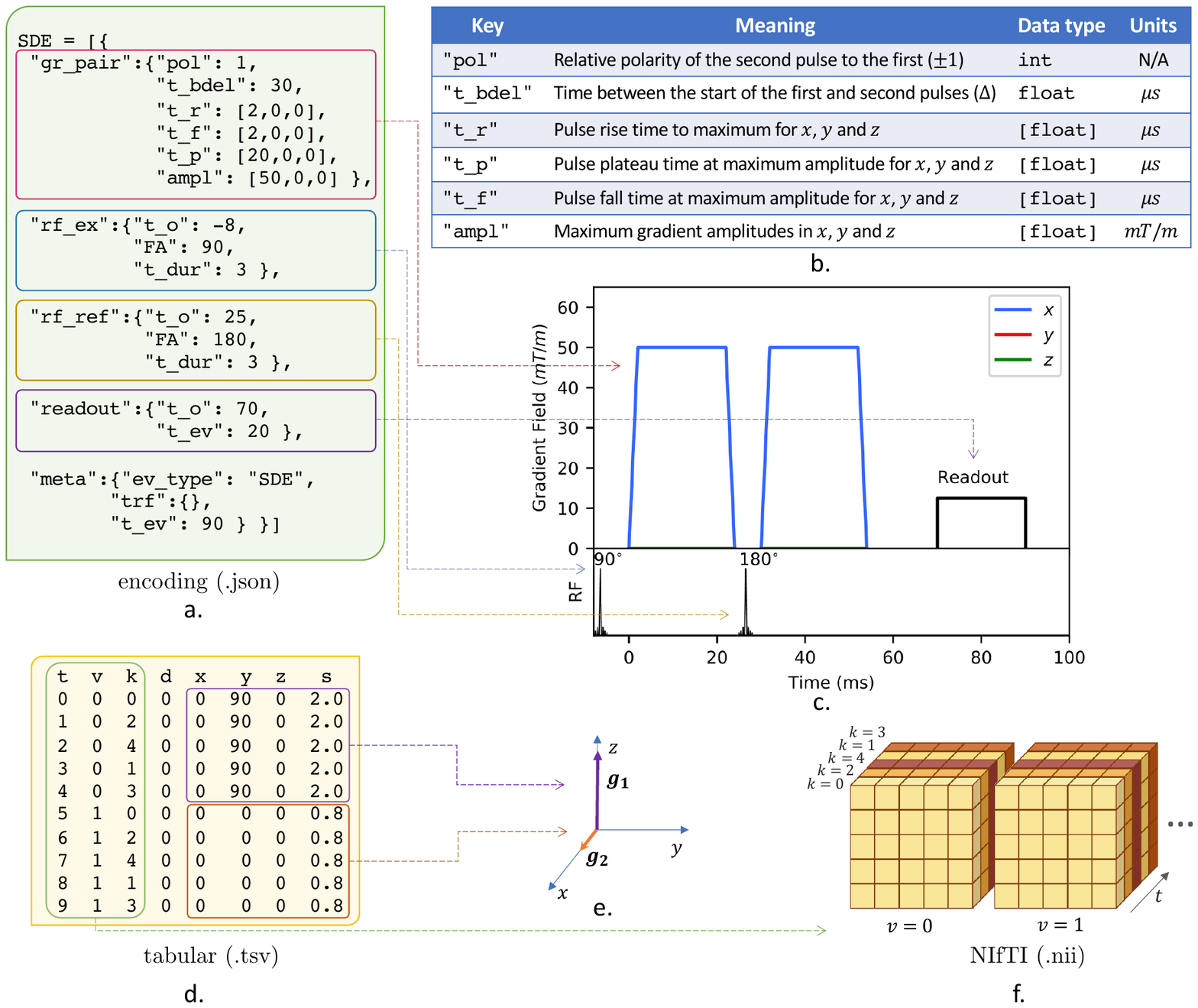}
    \caption{Encoding, tabular, and NIfTI files for a particular experiment based on the SDE sequence. The encoding file (a.) describes the sequence (c.) using subevents with keys defining the corresponding parameters (b.). This file is then related to the imaging data by the tabular file (d.), which allows not only to interpret the acquisition order (f.), but also the transformations of the base sequence for each slice and volume (e.). (See Section~\ref{sec:ex1stej} for more details).}
    \label{fig:sde}
\end{figure}

This variation of the NMR diffusion experiment remains popular and is supported by BIDS for very particular experimental setups. Here, we use the aDWI extension to show the versatility that can be gained even for this basic sequence. Fig.~\ref{fig:sde} shows an outline of the corresponding encoding, tabular, and NIfTI files and their relation. The encoding file is built employing the modular system of events and subevents discussed in Section~\ref{sec:encoding_file}. In particular, we outline an event that offers a terse description of a pair of identical gradient pulses alongside two RF pulses, given as three sub-events of the parent event (Fig.~\ref{fig:sde}a.--c.). The three subevents are labelled \verb|"gr_pair"|, \verb|"rf_ex"|, and \verb|"rf_ref"|, representing the gradient pulse pair, the 90$^{\circ}$ RF excitation pulse, and the 180$^{\circ}$ RF refocusing pulse, respectively. A \verb|"readout"| subevent is also included for illustrative purposes, as well as the \verb|"meta"| subevent describing event-wide properties.

Each subevent uses a syntax appropriate to its intended use. Both RF subevents use (in this case) identical schema, but occur with different flip angles (\verb|"FA"|) and time offsets (\verb|"t_o"|). The gradient subevent assumes (\verb|"t_o":0|), and hence is omitted. Timings for the gradient subevent are given in a conventional way, in terms of rise (\verb|"t_r"|), fall (\verb|"t_f"|) and plateau (\verb|"t_p"|) times~\cite{bernstein_handbook_2004}, as well as a maximal amplitude (\verb|"ampl"|). Gradient pulse amplitudes are given as an array of three values corresponding to maximal amplitudes in $x$, $y$ and $z$ global coordinates. We indicate the spacing of these pulses with \verb|"t_bdel"|, using the $\Delta$ convention~\cite{shemesh_conventions_2016}. We indicate the relative polarity of the paired pulses with \verb|"pol"|, with $1$ and $-1$ denoting identical and opposite polarity, respectively. 

The prototypical sequence conveyed by the encoding object is hence descriptive of the SDE experiment carried out in a single direction, gradient strength and shape, usually condensed in the b-value. The tabular file (Fig.~\ref{fig:sde}.d) communicates how this sequences varies between slices and volumes, orchestrating the imaging protocol. Fig.~\ref{fig:sde}.b shows how slices within a 4D dataset are mapped to indices $t$, $k$, and $v$ within the tabular file. In the presented example, slices are not acquired sequentially, and therefore the user requires these ordering parameters introduced in the aDWI in order to interpret the data. Index $d$ indicates which of multiple prototypical encoding objects is used in this slice (here just one is used, and could consequently have been omitted). 

The tabular file permutes the prototypical sequence (here varying between volumes $v=0$ and $v=1$) using functors, commutative abstract operations on a group. Some functors are given using multiple values. For example, \verb|x,y,z| represents a set of Euler angles through which the prototypical sequence is rotated. Similarly, the object may be linearly scaled with index \verb|s|. Consequently, looking in detail at Fig.~\ref{fig:sde}.d, we see that \verb|s| takes values $2.0$ and $0.8$, and \verb|y| equals $90$ or $0$. This means that, for $v=0$, we rotate the sequence by 90$^{\circ}$ about the $y$-axis, and scale by $2.0$, resulting in $g_1$ in Fig.~\ref{fig:sde}.e. For $v=1$, we do not rotate the sequence, and scale by $0.8$, giving $g_2$ in Fig.~\ref{fig:sde}.e. This descriptive syntax is based on and extends the abstract \verb|.bvec| file, allowing the labelling of more than just rotations.

\subsection{Example 2: double diffusion encoding}\label{sec:ex2DDE}

In the double diffusion encoding (DDE; also known as double pulse gradient) experiment, two gradient pulse pairs are used in the encoding period, each in an arbitrary direction~\cite{shemesh_conventions_2016, callaghan_translational_2011}. In an environment displaying restricted diffusion in locally anisotropic pores, the direction of this anisotropy compared with that of the applied pulses will affect signal losses. With an appropriate DDE experiment, the direction and extent of this anisotropy may be inferred~\cite{jespersen_orientationally_2013}. The use of multiple gradient directions across multiple acquisitions permits estimation of microstructural parameters not accessible with SDE, such as the $\mu$FA~\cite{shemesh_conventions_2016, callaghan_translational_2011, lasic_microanisotropy_2014} and comparable metrics~\cite{jespersen_orientationally_2013}.

\begin{figure}[h]
    \centering
    \includegraphics[scale = .8, trim=4.5cm 1cm 2cm 8cm, clip]{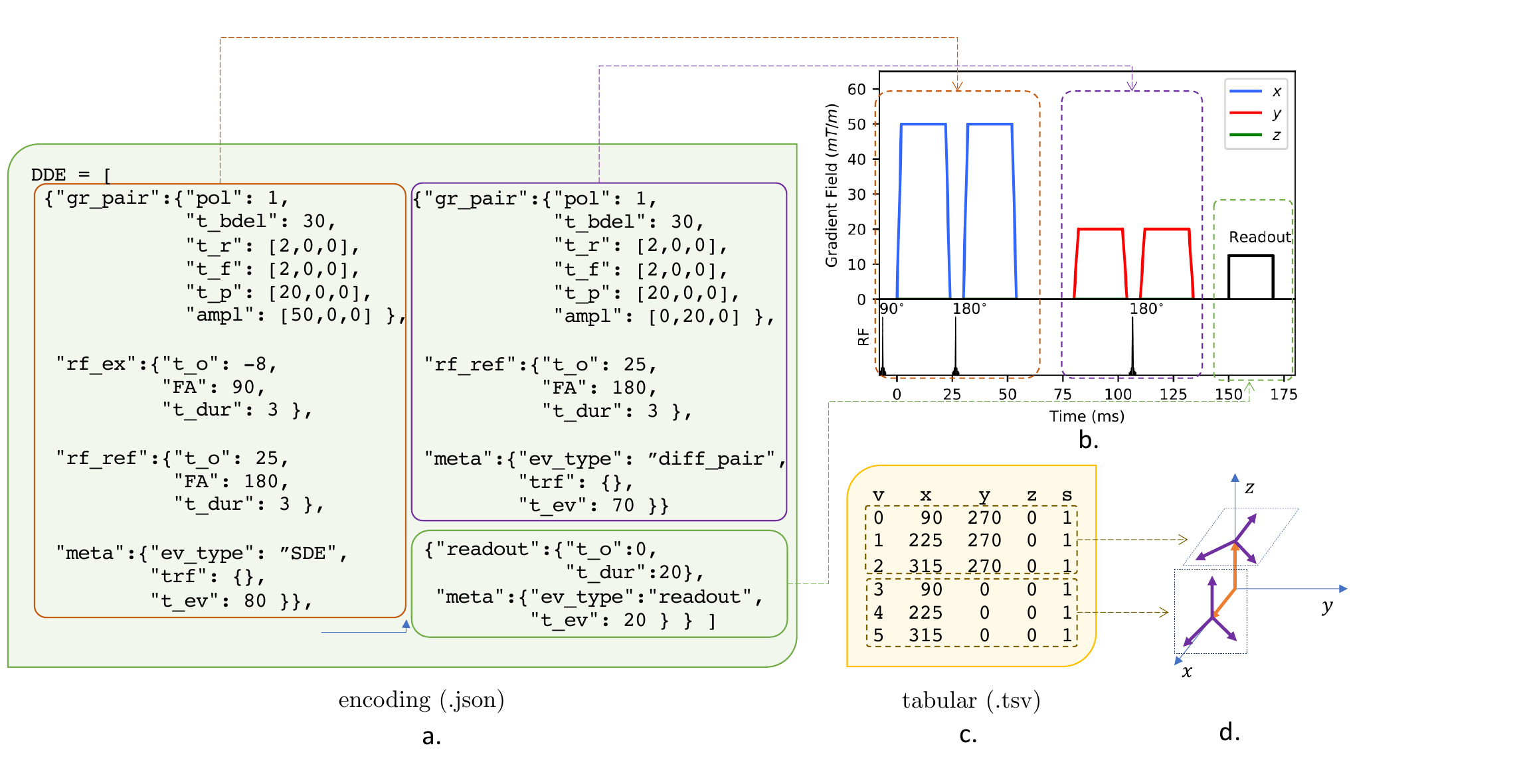}
    \caption{Example of a DDE experiment and its representation with the aDWI-BIDS. A single prototypical DDE sequence (b.) is described in the encoding file (a.), with successive rotations from different acquisitions indicated by the tabular file (c). Six orientational variations of the sequence are run, giving two diffusion wavevectors probed in the primary directions (d., in orange), with three secondary directions (d., in purple) for each primary direction probed. Rotations about the $x$, $y$ and $z$ axes of the prototypical sequence are given under their respective column headers in the tabular file, leading to six rows, one for each acquisition.}
    \label{fig:DDE}
\end{figure}

Fig.~\ref{fig:DDE} depicts the encoding and tabular files for a DDE experiment. The prototypical encoding object used in this example (Fig.~\ref{fig:DDE}.a) is more complex than that of the SDE sequence. As two pulse pairs are used, we employ two slightly different events: the first encapsulating both 90$^{\circ}$ and 180$^{\circ}$ RF pulses, while the second only incorporating a 180$^{\circ}$ RF pulse. This structure can logically be extended to multiple diffusion encodings by adding additional \verb|"diff_pair"| objects. Otherwise, the subevents used are identical to that used for SDE. Unlike in the previous sequence, readout is contained within its own event. Once again, the readout is given in a very loose schematic format - this could easily be replaced with a more complex and descriptive one (see Section~\ref{sec:RF_RO}).

Because two directions are interrogated simultaneously, they may be split into a first and second pulse, or alternatively a primary and a secondary axis. It is hence typical to vary only the secondary axis for several measurements, then adjust to a new primary axis, then repeat this variation of the secondary axis~\cite{jespersen_orientationally_2013}. This may be described simply in aDWI by employing a set of Euler angles rotating a prototypical encoding object, as shown in Figs.~\ref{fig:DDE}.c-d. Here, \verb|x, y| and \verb|z| are used to rotate the prototypical sequence as required, simplifying the sequence description to a very great extent. The relative intensity of the pulses goes unchanged for this particular use case (although it may have been modified). This example demonstrates that minimal repetition of data is needed to describe conceptually similar diffusion encoding, even when complex 3D rotations must be applied to relate parameter states. 

\subsection{Example 3: free gradient waveforms}\label{sec:fwf}

A notable development in diffusion imaging has been the utilisation of optimised, arbitrary gradient waveforms as diffusion encoding pulses~\cite{szczepankiewicz_quantification_2015, szczepankiewicz_gradient_2020}. The use of waveforms varying throughout the encoding results in, e.g., controllable sensitivity to microscopic metrics. This technique is implicitly broad; 'free' waveforms may be optimised to match many imaging and artifact criteria simultaneously, offering advantages in numerous arenas.

The use of free gradient waveforms (FGW) will normally result in a need for them to be known for pre-processing, model fitting, and artifact correction~\cite{szczepankiewicz_maxwell-compensated_2019}. This complication has been addressed with the sharing of tabular files, giving gradient amplitudes as a function of time. aDWI-BIDS includes support for this type of data in a dense binary format. 

\begin{figure}[h]
    \centering
    \includegraphics[scale = .8, trim=4.5cm 0cm 2cm 7cm, clip]{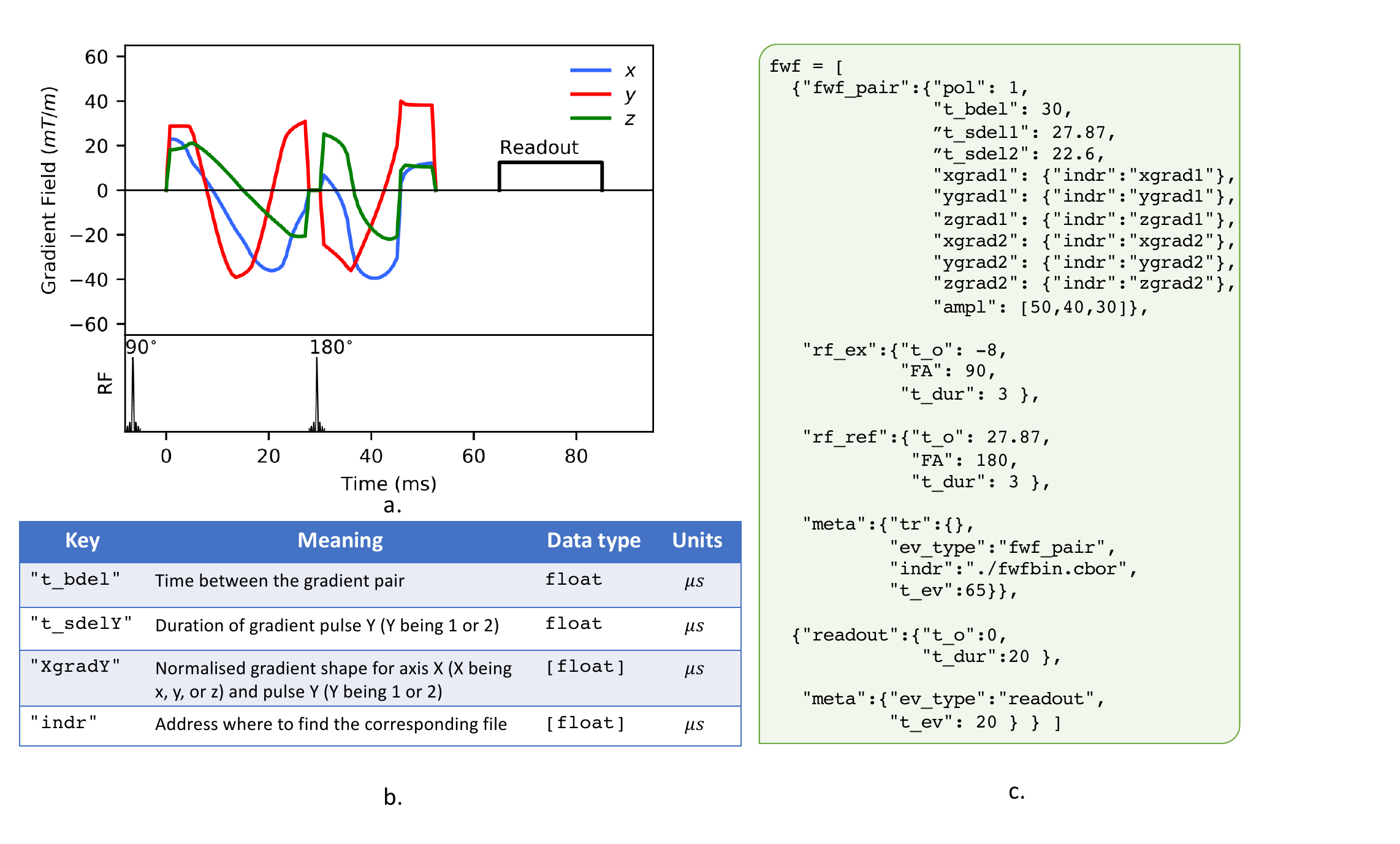}
    \caption{Free gradient waveform sequence description by means of the aDWI-BIDS. A prototypical sequence (a.) is conveyed by an encoding object (c.) using an indirected array of normalised amplitudes with equal spacing in time. Keys for the gradient subevent are given in b.}
    \label{fig:fwf}
\end{figure}

Fig.~\ref{fig:fwf} shows a spin-echo FGW sequence and its corresponding encoding file. As done in the SDE and DDE examples, standardised events are employed. In this case, due to multiple large arrays describing the gradient waveforms, we make extensive use of the indirection concept introduced in Section~\ref{sec:indr_perturb}. To maintain the readability of files, we save the gradient data in a separate file, \verb|fwfbin.cbor|, located in the same directory as the encoding file. This is a binary CBOR~\cite{CBOR} document, and hence is object-like.
Within the encoding file (Fig.~\ref{fig:fwf}.c), we use the key-value pair \verb|"indr":"value"|, implying that the file referenced by \verb|"meta":"indr"| should be accessed with the data referenced by key \verb|"indr"|, and the \verb|"indr":"value"| object replaced by the resultant accessed data. This causes this pattern to be treatable as a simple array, albeit in a dense binary format. Timings are provided with the same notation as for a trapezoidal pulse pair, though rise, fall and plateau times are rolled into a single variable per pulse ($\delta_1$ and $\delta_2$, as is standard notation in the field~\cite{szczepankiewicz_gradient_2020}). Pulse start times are separated by \verb|"t_bdel"| in the same fashion as for SDE. Amplitudes are normalised to unity, with their maximal value given for $x$, $y$ and $z$ in a 3-element \verb|"ampl"| array. This maximises the dynamic range available for this behaviour.

\subsection{Beyond diffusion encoding}\label{sec:RF_RO}
In addition to labelling of diffusion encoding, more general regions of sequence activity can be conveyed by aDWI-BIDS. We consider two cases: That of a complex readout, and that of a complex slice selection procedure.

\begin{figure}[h]
    \centering
    \includegraphics[scale = .8, trim=4.5cm 0cm 2cm 3.0cm, clip]{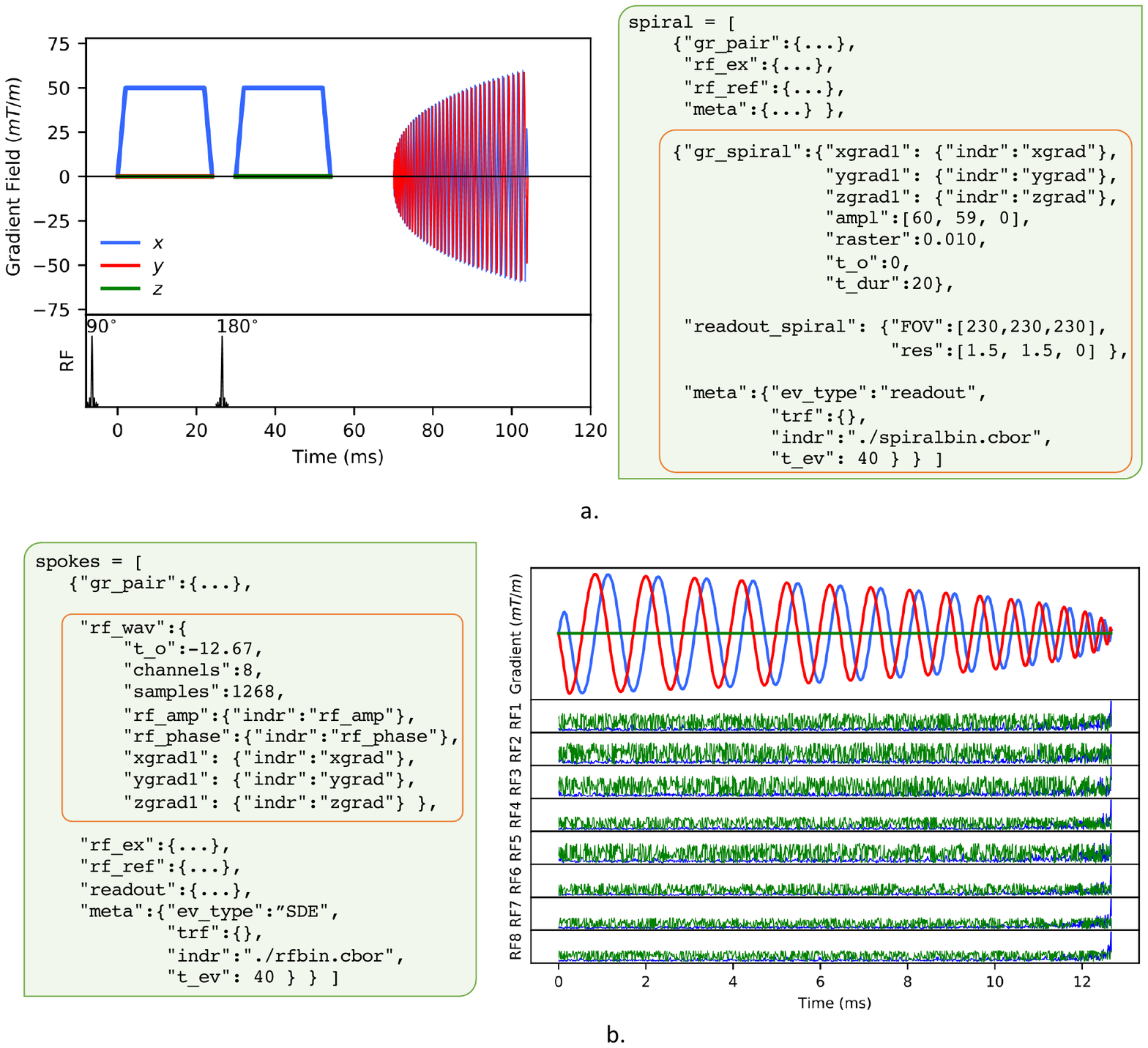}
    \caption{Spiral readout and inner volume excitation conveyed in aDWI-BIDS. a. Encoding object and sequence diagram pair for a spiral readout using indirected binary files. b. 8-channel RF slice selection routine, with related SS gradients. For RF1-8, green represents signal phase, and blue signal amplitude.}
    \label{fig:RF_RO}
\end{figure}

Spiral readout trajectories (SR) in diffusion MRI are appealing for several reasons: resilience to motion and and flow artefacts (vs. EPI), high SNR~\cite{lee_signal--noise_2021}, short TE~\cite{wilm_minimizing_2020}, and efficient use of gradient hardware~\cite{delattre_spiral_2010}. SR methods incur  complications: gradient waveforms are required, long readouts exacerbate inhomogeneity and concomitant field artefacts, and the processing spiral-acquired data requires a more complex approach than simple FFT. Furthermore, design of waveforms used for SR typically accounts for $B_0$ inhomogeneity unique to a scanner (leading to a unique waveform, potentially for each scanner / protocol). These issues flag a need for rich metadata, ideally detailing the waveform used exactly. Figure \ref{fig:RF_RO}.a shows such a waveform given as an indirected binary object alongside metadata.

Similarly, another technique which benefits from embedded waveforms is inner volume RF excitation approaches. Figure \ref{fig:RF_RO}.b shows a partial sequence to excite a semicircular volume in parallel. Eight RF waveforms are used alongside a three-axis gradient waveform, all given as binary indirected objects. Multiple channels are handled simply, and the metadata to interpret them is all present in the event. Data in this example is compressed from 315 kilobytes to just 167 kilobytes. The use of waveform compression techniques could lower this at little cost.

\newpage
\section{Discussion}\label{sec:discussion}

We have shown that aDWI-BIDS, as proposed, is a robust and generic means of communicating sequence activity under a wide variety of conditions. The structure presented is able to provide metadata used in analysis or postprocessing for numerous diffusion experiments. This includes support for arbitrary constructions of activity spaced in time, comprising potentially heterogenous mixes of waveform and parameterised activity. The proposed structure supports complex experiments in terms of parametric variation where possible, minimising repetition of data. These properties occur in the framework of a BIDS-compliant, schematically summarisable data standard.

Our proposal offers several attractive qualities for recording DWI metadata within BIDS. These are of merit to sequence designers in several specialisations, including diffusion encoding, readout and RF pulse methods. The data recorded displays the following properties:

\begin{itemize}
    \item Complete gradient and RF behaviour, given either parametrically or as values at discrete times.
    \item Modular construction; components may express different ontologies which are shared online.
    \item Dense representation of data and a means to reuse it.
    \item Multiple sequence behaviours described simultaneously, e.g., diffusion preparation, readout or RF pulses.
    \item Unique data given on a slice-by-slice or volume-by-volume basis.
    \item Derivative volumes included in multidimensional data are explicitly labelled (rather than inferred~\cite{li_first_2016}).
\end{itemize}

These characteristics make the aDWI-BIDS extension a comprehensive framework to share DWI experiments and processing pipelines, allowing easy replication of results and methods take-up.

\subsection{Comparison with other data standards}

In determining the needs of aDWI-BIDS, researchers from multiple institutions working with diffusion sequences were asked which metadata they wished for in an ideal standard, and how best to ensure the compatibility of this standard with existing methods. Several key requirements were flagged in areas where it was felt existing standards were lacking:
\begin{itemize}
    \item To retain backwards compatibility with FSL's \path{.bvec} and \path{.bval} files for describing b-values and b-vectors, respectively. 
    \item To record critical processing parameters like $\delta$ and $\Delta$ in a simple and accessible format.
    \item To provide support for multiple types of pulse shape, potentially in a single encoding period.
    \item To meet a long-term goal for a full sequence history, which was motivated by the appeal of amorphous pulse shapes and the need to provide them as metadata.
    \item Where possible, to offer parameterised versions of pulses to facilitate dense representation.
    \item To simply describe instances where a sequence is permuted in a parameterisable way (e.g., rotations).
\end{itemize}

Several existing approaches seek to codify the storage of DWI data for advanced sequences, being the most notable the eXperimental Parameter Structure~\cite{nilsson_open-source_2018} (XPS), the BIDS for DWI~\cite{gorgolewski_brain_2016}, and the DICOM standard. 


    


The XPS gives parameters volume-by-volume in a tabular format, and so is appropriate for single-volume-single-encoding acquisitions. Full specification is not given for how to represent gradient waveforms. Multiple diffusion encoding (MDE) is the broad focus of the standard, separating gradient pulses according to a collection of indexed $\delta_n$ and $\Delta_n$ values. XPS data is stored as a concentenated series of struct arrays, with fixed keys to identify parameters in a multiple diffusion encoding experiment. Some sequences will supply redundant data defined by this method, such as $t_m$ or $\Delta$, which are commonly fixed for acquisitions on the same shell. The XPS is reliant on proprietary software (MATLAB) to decode its struct arrays, limiting access to the standard. Overall, the XPS provides several commonly varied parameters in an MDE experiment, and was key in the determination of an ideal BIDS extension. Ultimately, its main limitation is its development in isolation to other medical imaging standards, making it difficult to embed into multimodal processing pipelines.

Comparatively, BIDS for DWI (not to be confused with BIDS-aDWI) is not able to describe multiple diffusion encoding, gradient waveforms, and like the XPS does not describe slice-level variation in diffusion encoding. BIDS for DWI is reliant on filename tags and its JSON sidecar to inform processing. It is typical to exploit \verb|.bvec| and \verb|.bval| files to imply experimental parameter sets. Alongside a clear sequence name in the sidecar, this approach can yield flexible processing behaviour when a single order of acquisition is used for all sequences using that name. This approach fails to supply parameter variation between volumes or slices in an acquisition, and hence tracking this variation is reliant again on inference, and secondary metadata files of heterogenous type. BIDS does, however, have a detailed process for extension, which we exploit with BIDS-aDWI. The advantages of belonging to the overarching BIDS is a crucial for maximising its adoption and maintenance, and our standard grows the capabilities of BIDS without compromising its flexibility. 

DICOM is perhaps the most widely supported and documented neuroimaging format~\cite{li_first_2016}. Its rigorous, language-agnostic definitions, dense data representation and certifiability win it much support from medical device manufacturers. Despite this, decades of use have resulted in large lexicons of key-concept correspondences, and adding to these requires substantial review. Conflict with past data is considered unacceptable, and major changes are always avoided. 

The data relayed by aDWI-BIDS is ultimately representable as a mix of JSON data and tabular data. Annex F of part 18 of the DICOM standard details how JSON data may be explicitly represented in DICOM. This specifies that objects must display a concrete representation, with JSON keys must be mapped to a DICOM tag. This requires that every possible key possesses a unique tag, requiring a substantial dictionary of tags to facilitate this. We ultimately consider representation within DICOM desirable, but stymied by potential conflicts between different tags, and the large number of tags needed to describe a broad collection of sequences. The use of private (or vendor) tags alleviates some of these concerns by splitting dictionaries by vendor, as well as bypassing the lengthy DICOM review process. Consequently, we consider eventual inclusion of diffusion metadata directly within a DICOM as a realistic goal.


\subsection{Design decisions in aDWI-BIDS}

The modular structure employed in aDWI-BIDS must overcome numerous common pitfalls in object-oriented design, such as object ambiguity, verbosity, complexity and efficiency. Overall, the standard overcomes these concerns with a 'flyweight' structure, where the prototypical object represents a 'flyweight' (i.e., a high inertia object) that is expensive to interpret, but is extensively reused to best benefit from its inertia. Its permutations, by comparison, are simple to apply to the prototypical object. The flyweight is hence optimised for density and flexibility, leading to the event-based structure actually used. 

The event structure permits significant latitude in design. The ability to describe overlapping blocks permits sequences to be written in terms of their components. Timing based on event locality (as opposed to a single common time axis) limits precision error accumulation and inherently wasteful long number representations. 

By adopting a JSON schema for all unique events, we can indicate the irreducible elements of an MRI sequence in an extensible, but unambiguous way. Grouping of events into containers permits identical keys with readable names. Furthermore, it permits simple heuristic determination of sequence type by comparison of a JSON against an appropriate schema. For this reason, MR scientists may simply generate JSON schemas specifying their sequences, which will serve as a basis for their accurate representation. We provide schemas for the sequences treated in this manuscript (SDE, DDE, and FGW), as well as others (e.g., oscillating gradients~\cite{baron_oscillating_2014} and FEXI~\cite{nilsson_noninvasive_2013}) in the aDWI repository ( \url{https://github.com/JAgho/aDWI-BIDS}).

\subsection{aDWI-BIDS and choice}

The structure described is a framework designed to be expanded upon in a modular way. The event schemas given as examples are merely demonstrations of how the components of a system may be used. Thoughtful design may find certain structures easier to modify, or apply parametric transformations to. Where this is unnecessary, simple structures built from generic events work as expected, with the only cost being larger data.

Similarly, designing for concision is a matter left to developers. For this reason, binary indirections are given as a optional tool to lighten otherwise storage heavy arrays of data. The choice to use CBOR~\cite{CBOR} is by no means final, though a binary structured object acting similarly to JSON is required to best leverage the flyweight design used. Comparable data standards include MessagePack~\cite{messagepack}, BSON~\cite{bson} and UBJSON~\cite{UBJSON}. 


The tabular file itself presents significant challenges in terms of choice. Consider a simple DDE sequence, run with varying $\delta$. This will imply the movement of not only gradient pulse pairs, but also of related RF activity. Thoughtful design will permit a single parameter $\delta$ to act as a transformation of the prototypical object by grouping related subevents together and defining the explicit consequences of a transformation. Careless design still permits this, but the tabular file may have to act individually on subevents within the encoding object, for a less concise file. Consequently, to ensure events and code to process them are universally available, we suggest a single repository for event and subevent schemata, alongside transformations. These can be intelligently implemented as programmatic classes, with transformations, expansions, etc., given as methods of the object.

\subsection{aDWI-BIDS prospects}

The recording of archive-quality data is of great importance in many applications, machine learning being amongst the most prominent~\cite{folmsbee_fragile_2019}. Central to this aim is access to large datasets describing basal populations in detail. While large and robust datasets of DTI data already exist, the superior detail provided by microstructure-resolving sequences is not represented in this data.

The adoption of the proposed extension would allow scientists to share not only data in a comprehensive way but, as importantly, software designed with the extension as a basis. BIDS data ideally contains sufficient information to determine the correct processing method to use for a given combination of parameters. This is modularised in the form of BIDS Apps~\cite{gorgolewski_bids_2017}, container instances representing whole or sections of processing pipelines. These are controlled using a uniform syntax, permitting the construction of automated processing systems, which may operate on multiple modalities in a dataset. BIDS apps are supplied as both Docker and Singularity containers, facilitating small- and large-scale processing of datasets (e.g.,~\cite{smith_resting-state_2013, glasser_minimal_2013}, or \cite{esteban_fmriprep_2019}).

As a demonstration of a simple BIDS-App, we generated an automated plotter to operate on aDWI-BIDS data. MISP-plot (\url{https://github.com/JAgho/MISP_plot}) is a Python tool that leverages Matplotlib to generate sequence diagrams based on encoding objects. Indeed, every sequence diagram used in this paper was generated programmatically based on the data supplied in the objects shown. All waveforms, amplitude, durations and indirections are available with a simple access pattern, which is demonstrated in the repository. With corresponding tabular data, unique diagrams may be generated for every instance of an experiment.

We wish for aDWI-BIDS to support detailed heuristics, aiding automated processing of data collected with multiple DWI modalities. This would permit, for example, detailed pipeline comparison studies, or population studies on otherwise difficult to compare data. The given structure may be be matched directly against a set of JSON schemas to identify type and purpose, and then compared against known sequences to identify a particular processing approach to employ.

Open data standards are crucial for making scientific progress, allowing researchers to normalise the language used and consequently promoting cross-fertilisation between fields. Such standards have formed the basis of extremely successful collaborations where coordination was shown to play a key role, e.g., for the detection of gravitational waves, or management of astronomical equipment. These concepts are less developed in uncoordinated fields dominated by organic growth such as medical imaging, leading to a myriad of nomenclatures that hinder transparent data and methods sharing. In this context, the BIDS was presented as an overarching framework to reconcile differences in nomenclature, establishing requirements for accurate and comprehensive data sharing. By building upon the BIDS, we expect the community to join forces towards an ultimate goal of obtaining better estimates of tissue properties, in vivo and non-invasively, enabling big data studies from heterogeneous sources as never before.

\section*{Acknowledgements}
This work was supported by the Science and Technology Facilities Council, UK, through grants ST/00209X/1 (MIDaC) and Impact Acceleration Account (MISP, Cardiff University).

Derek Jones was supported by a Wellcome Investigator Award (to DKJ; Grant reference: 096646/Z/11/Z, and by a Wellcome Trust Strategic Award Grant reference: 104943/Z/14/Z)

CMWT was supported by a Sir Henry Wellcome Fellowship (215944/Z/19/Z) and a Veni grant (17331) from the Dutch Research Council (NWO)



\bibliographystyle{ieeetr}  
\bibliography{references}  

\begin{thebibliography}{10}

\bibitem{assaf_composite_2005}
Y.~Assaf and P.~J. Basser, ``Composite hindered and restricted model of
  diffusion ({CHARMED}) {MR} imaging of the human brain,'' {\em NeuroImage},
  vol.~27, pp.~48--58, Aug. 2005.

\bibitem{zhang_noddi_2012}
H.~Zhang, T.~Schneider, C.~A. Wheeler-Kingshott, and D.~C. Alexander,
  ``{NODDI}: {Practical} in vivo neurite orientation dispersion and density
  imaging of the human brain,'' {\em NeuroImage}, vol.~61, pp.~1000--1016, July
  2012.

\bibitem{cory_measurement_1990}
D.~G. Cory, ``Measurement of translational displacement probabilities by {NMR}:
  {An} indicator of compartmentation,'' {\em Magnetic Resonance in Medicine},
  vol.~14, no.~3, pp.~435--444, 1990.
\newblock \_eprint:
  https://onlinelibrary.wiley.com/doi/pdf/10.1002/mrm.1910140303.

\bibitem{shemesh_conventions_2016}
N.~Shemesh, S.~N. Jespersen, D.~C. Alexander, Y.~Cohen, I.~Drobnjak, T.~B.
  Dyrby, J.~Finsterbusch, M.~A. Koch, T.~Kuder, F.~Laun, M.~Lawrenz,
  H.~Lundell, P.~P. Mitra, M.~Nilsson, E.~Özarslan, D.~Topgaard, and C.-F.
  Westin, ``Conventions and nomenclature for double diffusion encoding {NMR}
  and {MRI},'' {\em Magnetic Resonance in Medicine}, vol.~75, no.~1,
  pp.~82--87, 2016.

\bibitem{ozarslan_microscopic_2008}
E.~Özarslan and P.~J. Basser, ``Microscopic anisotropy revealed by {NMR}
  double pulsed field gradient experiments with arbitrary timing parameters,''
  {\em The Journal of Chemical Physics}, vol.~128, p.~154511, Apr. 2008.
\newblock Publisher: American Institute of Physics.

\bibitem{callaghan_komlosh_2002}
P.~Callaghan and M.~Komlosh, ``Locally anisotropic motion in a macroscopically
  isotropic system: {Displacement} correlations measured using double pulsed
  gradient spin-echo {NMR},'' {\em Magnetic Resonance in Chemistry}, vol.~40,
  no.~SPEC. ISS., pp.~S15--S19, 2002.

\bibitem{ozarslan_mr_2007}
E.~Özarslan and P.~J. Basser, ``{MR} diffusion–“diffraction” phenomenon
  in multi-pulse-field-gradient experiments,'' {\em Journal of Magnetic
  Resonance}, vol.~188, pp.~285--294, Oct. 2007.

\bibitem{clark_diffusion_2001}
C.~Clark, M.~Hedehus, and M.~Moseley, ``Diffusion time dependence of the
  apparent diffusion tensor in healthy human brain and white matter disease,''
  {\em Magnetic Resonance in Medicine}, vol.~45, no.~6, pp.~1126--1129, 2001.

\bibitem{shemesh_measuring_2009}
N.~Shemesh, E.~Özarslan, P.~J. Basser, and Y.~Cohen, ``Measuring small
  compartmental dimensions with low-q angular double-{PGSE} {NMR}: {The} effect
  of experimental parameters on signal decay,'' {\em Journal of Magnetic
  Resonance}, vol.~198, pp.~15--23, May 2009.

\bibitem{descoteaux_hardi_2015}
M.~Descoteaux, ``High {Angular} {Resolution} {Diffusion} {Imaging} ({HARDI}),''
  in {\em Wiley {Encyclopedia} of {Electrical} and {Electronics}
  {Engineering}}, pp.~1--25, American Cancer Society, 2015.

\bibitem{avram_vivo_2013}
A.~V. Avram, E.~Özarslan, J.~E. Sarlls, and P.~J. Basser, ``In vivo detection
  of microscopic anisotropy using quadruple pulsed-field gradient ({qPFG})
  diffusion {MRI} on a clinical scanner,'' {\em NeuroImage}, vol.~64,
  pp.~229--239, Jan. 2013.

\bibitem{topgaard_isotropic_2015}
D.~Topgaard, ``Isotropic diffusion weighting using a triple-stimulated echo
  pulse sequence with bipolar gradient pulse pairs,'' {\em Microporous and
  Mesoporous Materials}, vol.~205, pp.~48--51, Mar. 2015.

\bibitem{topgaard_multidimensional_2017}
D.~Topgaard, ``Multidimensional diffusion {MRI},'' {\em Journal of Magnetic
  Resonance}, vol.~275, pp.~98--113, Feb. 2017.

\bibitem{westin_q-space_2016}
C.-F. Westin, H.~Knutsson, O.~Pasternak, F.~Szczepankiewicz, E.~Özarslan,
  D.~van Westen, C.~Mattisson, M.~Bogren, L.~J. O'Donnell, M.~Kubicki,
  D.~Topgaard, and M.~Nilsson, ``Q-space trajectory imaging for
  multidimensional diffusion {MRI} of the human brain,'' {\em NeuroImage},
  vol.~135, pp.~345--362, July 2016.

\bibitem{sjolund_constrained_2015}
J.~Sjölund, F.~Szczepankiewicz, M.~Nilsson, D.~Topgaard, C.-F. Westin, and
  H.~Knutsson, ``Constrained optimization of gradient waveforms for generalized
  diffusion encoding,'' {\em Journal of Magnetic Resonance}, vol.~261,
  pp.~157--168, Dec. 2015.

\bibitem{szczepankiewicz_quantification_2015}
F.~Szczepankiewicz, S.~Lasič, D.~van Westen, P.~C. Sundgren, E.~Englund, C.-F.
  Westin, F.~Ståhlberg, J.~Lätt, D.~Topgaard, and M.~Nilsson,
  ``Quantification of microscopic diffusion anisotropy disentangles effects of
  orientation dispersion from microstructure: applications in healthy
  volunteers and in brain tumors,'' {\em NeuroImage}, vol.~104, pp.~241--252,
  Jan. 2015.

\bibitem{HUTTER_2018214}
J.~Hutter, D.~J. Christiaens, T.~Schneider, L.~Cordero-Grande, P.~J. Slator,
  M.~Deprez, A.~N. Price, J.-D. Tournier, M.~Rutherford, and J.~V. Hajnal,
  ``Slice-level diffusion encoding for motion and distortion correction,'' {\em
  Medical Image Analysis}, vol.~48, pp.~214 -- 229, 2018.

\bibitem{hutter_dynamic_2017}
J.~Hutter, D.~Christiaens, M.~Deprez, L.~Cordero-Grande, P.~Slator, A.~Price,
  M.~Rutherford, and J.~V. Hajnal, ``Dynamic {Field} {Mapping} and {Motion}
  {Correction} {Using} {Interleaved} {Double} {Spin}-{Echo} {Diffusion}
  {MRI},'' in {\em Medical {Image} {Computing} and {Computer} {Assisted}
  {Intervention} - {MICCAI} 2017} (M.~Descoteaux, L.~Maier-Hein, A.~Franz,
  P.~Jannin, D.~L. Collins, and S.~Duchesne, eds.), Lecture {Notes} in
  {Computer} {Science}, (Cham), pp.~523--531, Springer International
  Publishing, 2017.

\bibitem{gal_data_2019}
M.~S. Gal and D.~L. Rubinfeld, ``Data {Standardization} {Symposium},'' {\em New
  York University Law Review}, vol.~94, no.~4, pp.~737--770, 2019.

\bibitem{data_harmon_Bittner2021}
M.-I. Bittner, ``Rethinking data and metadata in the age of machine
  intelligence,'' {\em Patterns}, vol.~2, p.~100208, Feb. 2021.

\bibitem{gorgolewski_brain_2016}
K.~J. Gorgolewski, T.~Auer, V.~D. Calhoun, R.~C. Craddock, S.~Das, E.~P. Duff,
  G.~Flandin, S.~S. Ghosh, T.~Glatard, Y.~O. Halchenko, D.~A. Handwerker,
  M.~Hanke, D.~Keator, X.~Li, Z.~Michael, C.~Maumet, B.~N. Nichols, T.~E.
  Nichols, J.~Pellman, J.-B. Poline, A.~Rokem, G.~Schaefer, V.~Sochat,
  W.~Triplett, J.~A. Turner, G.~Varoquaux, and R.~A. Poldrack, ``The brain
  imaging data structure, a format for organizing and describing outputs of
  neuroimaging experiments,'' {\em Scientific Data}, vol.~3, pp.~1--9, June
  2016.

\bibitem{lopez-novoa_bids_2019}
U.~Lopez-Novoa, C.~Charron, J.~Evans, and L.~Beltrachini, ``The {BIDS}
  {Toolbox}: {A} web {Service} to {Manage} {Brain} {Imaging} {Datasets},'' in
  {\em 2019 {IEEE} {SmartWorld}, {Ubiquitous} {Intelligence} {Computing},
  {Advanced} {Trusted} {Computing}, {Scalable} {Computing} {Communications},
  {Cloud} {Big} {Data} {Computing}, {Internet} of {People} and {Smart} {City}
  {Innovation} ({SmartWorld}/{SCALCOM}/{UIC}/{ATC}/{CBDCom}/{IOP}/{SCI})},
  pp.~378--382, Aug. 2019.

\bibitem{BIDS_PET}
M.~Ganz, ``{BIDS} {BEP} 009: Positron emission tomography.''
\newblock https://github.com/bids-standard/bids-specification/pull/633.

\bibitem{pernet_eeg-bids_2019}
C.~R. Pernet, S.~Appelhoff, K.~J. Gorgolewski, G.~Flandin, C.~Phillips,
  A.~Delorme, and R.~Oostenveld, ``{EEG}-{BIDS}, an extension to the brain
  imaging data structure for electroencephalography,'' vol.~6, no.~1, p.~103,
  2019.

\bibitem{niso_meg-bids_2018}
G.~Niso, K.~J. Gorgolewski, E.~Bock, T.~L. Brooks, G.~Flandin, A.~Gramfort,
  R.~N. Henson, M.~Jas, V.~Litvak, J.~T.~Moreau, R.~Oostenveld, J.-M.
  Schoffelen, F.~Tadel, J.~Wexler, and S.~Baillet, ``{MEG}-{BIDS}, the brain
  imaging data structure extended to magnetoencephalography,'' vol.~5, no.~1,
  p.~180110, 2018.

\bibitem{bernstein_handbook_2004}
M.~A. Bernstein, {\em Handbook of {MRI} pulse sequences}.
\newblock Elsevier Academic Press, 2004.

\bibitem{baron_oscillating_2014}
C.~A. Baron and C.~Beaulieu, ``Oscillating gradient spin-echo ({OGSE})
  diffusion tensor imaging of the human brain,'' {\em Magnetic Resonance in
  Medicine}, vol.~72, no.~3, pp.~726--736, 2014.

\bibitem{edelstein_making_2002}
W.~A. Edelstein, R.~A. Hedeen, R.~P. Mallozzi, S.-A. El-Hamamsy, R.~A.
  Ackermann, and T.~J. Havens, ``Making {MRI} {Quieter},'' {\em Magnetic
  Resonance Imaging}, vol.~20, pp.~155--163, Feb. 2002.

\bibitem{nilsson_noninvasive_2013}
M.~Nilsson, J.~Lätt, D.~van Westen, S.~Brockstedt, S.~Lasič, F.~Ståhlberg,
  and D.~Topgaard, ``Noninvasive mapping of water diffusional exchange in the
  human brain using filter-exchange imaging,'' {\em Magnetic Resonance in
  Medicine}, vol.~69, pp.~1572--1580, June 2013.

\bibitem{matthaei_steam_1986}
D.~Matthaei, J.~Frahm, A.~Haase, K.~D. Merboldt, and W.~Hänicke,
  ``Multipurpose {NMR} imaging using stimulated echoes,'' {\em Magnetic
  Resonance in Medicine}, vol.~3, no.~4, pp.~554--561, 1986.
\newblock \_eprint:
  https://onlinelibrary.wiley.com/doi/pdf/10.1002/mrm.1910030409.

\bibitem{merboldt_steam_1991}
K.~D. Merboldt, W.~Hänicke, and J.~Frahm, ``Diffusion imaging using stimulated
  echoes,'' {\em Magnetic Resonance in Medicine}, vol.~19, pp.~233--239, June
  1991.

\bibitem{yoshida_image_2016}
T.~Yoshida, A.~Urikura, K.~Shirata, Y.~Nakaya, S.~Terashima, and Y.~Hosokawa,
  ``Image quality assessment of single-shot turbo spin echo diffusion-weighted
  imaging with parallel imaging technique: a phantom study,'' {\em The British
  Journal of Radiology}, vol.~89, p.~20160512, July 2016.
\newblock Publisher: The British Institute of Radiology.

\bibitem{mjenkinson_nifti-1_nodate}
M.~Jenkinson, ``{NIfTI}-1 data format — neuroimaging informatics technology
  initiative,''
\newblock https://nifti.nimh.nih.gov/pub/dist/src/niftilib/nifti1.h.

\bibitem{handrews-json-schema-02}
A.~Wright, H.~Andrews, B.~Hutton, and G.~Dennis, ``{JSON Schema: A Media Type
  for Describing JSON Documents},'' internet-draft, Internet Engineering Task
  Force, Sept. 2019.
\newblock Work in Progress.

\bibitem{CBOR}
C.~Bormann and P.~Hoffman, ``Concise binary object representation ({CBOR}),''
  tech. rep., Dec. 2020.

\bibitem{Stejskal}
E.~O. Stejskal and J.~E. Tanner, ``Spin {Diffusion} {Measurements}: {Spin}
  {Echoes} in the {Presence} of a {Time}‐{Dependent} {Field} {Gradient},''
  {\em The Journal of Chemical Physics}, vol.~42, pp.~288--292, Jan. 1965.
\newblock Publisher: American Institute of Physics.

\bibitem{callaghan_translational_2011}
P.~T. Callaghan, {\em Translational {Dynamics} and {Magnetic} {Resonance}:
  {Principles} of {Pulsed} {Gradient} {Spin} {Echo} {NMR}}.
\newblock Oxford University Press, 1992.
\newblock Publication Title: Translational Dynamics and Magnetic Resonance.

\bibitem{jespersen_orientationally_2013}
S.~N. Jespersen, H.~Lundell, C.~K. Sønderby, and T.~B. Dyrby,
  ``Orientationally invariant metrics of apparent compartment eccentricity from
  double pulsed field gradient diffusion experiments,'' {\em NMR in
  Biomedicine}, vol.~26, no.~12, pp.~1647--1662, 2013.

\bibitem{lasic_microanisotropy_2014}
S.~Lasič, F.~Szczepankiewicz, S.~Eriksson, M.~Nilsson, and D.~Topgaard,
  ``Microanisotropy imaging: quantification of microscopic diffusion anisotropy
  and orientational order parameter by diffusion {MRI} with magic-angle
  spinning of the q-vector,'' {\em Frontiers in Physics}, vol.~2, 2014.
\newblock Publisher: Frontiers.

\bibitem{szczepankiewicz_gradient_2020}
F.~Szczepankiewicz, C.-F. Westin, and M.~Nilsson, ``Gradient waveform design
  for tensor-valued encoding in diffusion {MRI},'' {\em Journal of Neuroscience
  Methods}, p.~109007, Nov. 2020.

\bibitem{szczepankiewicz_maxwell-compensated_2019}
F.~Szczepankiewicz, C.-F. Westin, and M.~Nilsson, ``Maxwell-compensated design
  of asymmetric gradient waveforms for tensor-valued diffusion encoding,'' {\em
  Magnetic Resonance in Medicine}, vol.~82, no.~4, pp.~1424--1437, 2019.
\newblock \_eprint: https://onlinelibrary.wiley.com/doi/pdf/10.1002/mrm.27828.

\bibitem{lee_signal--noise_2021}
Y.~Lee, B.~J. Wilm, D.~O. Brunner, S.~Gross, T.~Schmid, Z.~Nagy, and K.~P.
  Pruessmann, ``On the signal-to-noise ratio benefit of spiral acquisition in
  diffusion {MRI},'' {\em Magnetic Resonance in Medicine}, vol.~85,
  pp.~1924--1937, Apr. 2021.

\bibitem{wilm_minimizing_2020}
B.~J. Wilm, F.~Hennel, M.~B. Roesler, M.~Weiger, and K.~P. Pruessmann,
  ``Minimizing the echo time in diffusion imaging using spiral readouts and a
  head gradient system,'' {\em Magnetic Resonance in Medicine}, vol.~84, no.~6,
  pp.~3117--3127, 2020.

\bibitem{delattre_spiral_2010}
B.~M.~A. Delattre, R.~M. Heidemann, L.~A. Crowe, J.-P. Vallée, and J.-N.
  Hyacinthe, ``Spiral demystified,'' {\em Magnetic Resonance Imaging}, vol.~28,
  pp.~862--881, July 2010.

\bibitem{li_first_2016}
X.~Li, P.~S. Morgan, J.~Ashburner, J.~Smith, and C.~Rorden, ``The first step
  for neuroimaging data analysis: {DICOM} to {NIfTI} conversion,'' {\em Journal
  of Neuroscience Methods}, vol.~264, pp.~47--56, May 2016.

\bibitem{nilsson_open-source_2018}
M.~Nilsson, F.~Szczepankiewicz, B.~Lampinen, A.~Ahlgren, J.~P.
  de~Almeida~Martins, S.~Lasic, C.-F. Westin, and D.~Topgaard, ``An open-source
  framework for analysis of multidimensional diffusion {MRI} data implemented
  in {MATLAB},'' 5355, (Paris Expo, Porte de Versailles, Paris), June 2018.

\bibitem{messagepack}
``{MessagePack} - an extremely efficient object serialization library.''
\newblock https://msgpack.org/index.html.

\bibitem{bson}
``{BSON} ({Binary} {JSON}) - a binary-encoded serialization of json-like
  documents.''
\newblock http://bsonspec.org/.

\bibitem{UBJSON}
``Universal {Binary} {JSON} {Specification} – {The} universally compatible
  format specification for {Binary} {JSON}..''
\newblock https://ubjson.org/.

\bibitem{folmsbee_fragile_2019}
J.~Folmsbee, S.~Johnson, X.~Liu, M.~Brandwein-Weber, and S.~Doyle, ``Fragile
  neural networks: the importance of image standardization for deep learning in
  digital pathology,'' in {\em Medical {Imaging} 2019: {Digital} {Pathology}},
  vol.~10956, p.~1095613, International Society for Optics and Photonics, Mar.
  2019.

\bibitem{gorgolewski_bids_2017}
K.~J. Gorgolewski, F.~Alfaro-Almagro, T.~Auer, P.~Bellec, M.~Capotă, M.~M.
  Chakravarty, N.~W. Churchill, A.~L. Cohen, R.~C. Craddock, G.~A. Devenyi,
  A.~Eklund, O.~Esteban, G.~Flandin, S.~S. Ghosh, J.~S. Guntupalli,
  M.~Jenkinson, A.~Keshavan, G.~Kiar, F.~Liem, P.~R. Raamana, D.~Raffelt, C.~J.
  Steele, P.-O. Quirion, R.~E. Smith, S.~C. Strother, G.~Varoquaux, Y.~Wang,
  T.~Yarkoni, and R.~A. Poldrack, ``{BIDS} apps: {Improving} ease of use,
  accessibility, and reproducibility of neuroimaging data analysis methods,''
  {\em PLOS Computational Biology}, vol.~13, p.~e1005209, Mar. 2017.
\newblock Publisher: Public Library of Science.

\bibitem{smith_resting-state_2013}
S.~M. Smith, C.~F. Beckmann, J.~Andersson, E.~J. Auerbach, J.~Bijsterbosch,
  G.~Douaud, E.~Duff, D.~A. Feinberg, L.~Griffanti, M.~P. Harms, M.~Kelly,
  T.~Laumann, K.~L. Miller, S.~Moeller, S.~Petersen, J.~Power,
  G.~Salimi-Khorshidi, A.~Z. Snyder, A.~T. Vu, M.~W. Woolrich, J.~Xu,
  E.~Yacoub, K.~Uğurbil, D.~C. Van~Essen, M.~F. Glasser, and {WU-Minn HCP
  Consortium}, ``Resting-state {fMRI} in the {Human} {Connectome} {Project},''
  {\em NeuroImage}, vol.~80, pp.~144--168, Oct. 2013.

\bibitem{glasser_minimal_2013}
M.~F. Glasser, S.~N. Sotiropoulos, J.~A. Wilson, T.~S. Coalson, B.~Fischl,
  J.~L. Andersson, J.~Xu, S.~Jbabdi, M.~Webster, J.~R. Polimeni, D.~C.
  Van~Essen, M.~Jenkinson, and {WU-Minn HCP Consortium}, ``The minimal
  preprocessing pipelines for the {Human} {Connectome} {Project},'' {\em
  NeuroImage}, vol.~80, pp.~105--124, Oct. 2013.

\bibitem{esteban_fmriprep_2019}
O.~Esteban, C.~J. Markiewicz, R.~W. Blair, C.~A. Moodie, A.~I. Isik,
  A.~Erramuzpe, J.~D. Kent, M.~Goncalves, E.~DuPre, M.~Snyder, H.~Oya, S.~S.
  Ghosh, J.~Wright, J.~Durnez, R.~A. Poldrack, and K.~J. Gorgolewski,
  ``{fMRIPrep}: a robust preprocessing pipeline for functional {MRI},'' {\em
  Nature Methods}, vol.~16, pp.~111--116, Jan. 2019.

\end{thebibliography}



\end{document}